\documentclass[final,5p,times,twocolumn]{elsarticle}

\usepackage{amssymb}
\usepackage{amsmath}
\biboptions{numbers,sort&compress}
\usepackage{xcolor}

\journal{Diamond and Related Materials}

\begin{document}

\begin{frontmatter}


\title{First-principles study of the formation of Group-IV--vacancy color centers in diamond}
\author[label1]{Xiangru Han}
\author[label1,label2,label3]{Adam Gali\corref{cor1}}
\ead{gali.adam@wigner.hun-ren.hu}
\cortext[cor1]{Corresponding author}

\affiliation[label1]{organization={HUN-REN Wigner Research Centre for Physics},
            addressline={P.O. Box 49},
             city={Budapest},
             postcode={H-1525},
             country={Hungary}}

\affiliation[label2]{organization={Department of Atomic Physics, Institute of Physics, Budapest University of Technology and Economics},
             addressline={M\H{u}egyetem rakpart 3},
             city={Budapest},
             postcode={H-1111},
            country={Hungary}}
            
\affiliation[label3]{organization={MTA--WFK Lend{\"u}let ``Momentum'' Semiconductor Nanostructures Research Group},
             addressline={P.O. Box 49},
             city={Budapest},
             postcode={H-1525},
            country={Hungary}}


\begin{abstract}
$XV$ ($X$ is Si, Ge, Sn, or Pb, and $V$ is the carbon vacancy) color centers are promising platforms for quantum computing and quantum information science due to their favorable optical and spin properties. They are commonly fabricated by ion implantation, which inevitably introduces vacancies and carbon self-interstitials (C$_i$) into the lattice. However, the microscopic formation mechanism remains unclear, limiting the yield and stability of $XV$ centers. Here, first-principles calculations are conducted to systematically investigate the roles of $V$ and C$_i$ in $XV$ formation. We find that $XV$ formation is energetically favorable, with large binding energies and low diffusion barriers. Nevertheless, $XV$ readily captures C$_i$ to form substitutional defect $X_s$, or captures an additional $V$ to form the complex $XV_2$. Charge transition levels show that oppositely charged reactants never meet in these reactions, and that Coulomb repulsion blocks them only when both reactants carry charges of the same sign, which for the capture of C$_i$ requires strongly n- or p-type conditions. These competing defect reactions destabilize the split-vacancy configuration and could account for the low formation yield of $XV$ centers, so that they appear at concentrations well below those expected from the implanted dose. Our results elucidate the microscopic origin of this intrinsic instability and suggest that Fermi-level control may suppress defect conversions and stabilize the split-vacancy configuration. 

\end{abstract}
\begin{graphicalabstract}
\includegraphics[width=\linewidth]{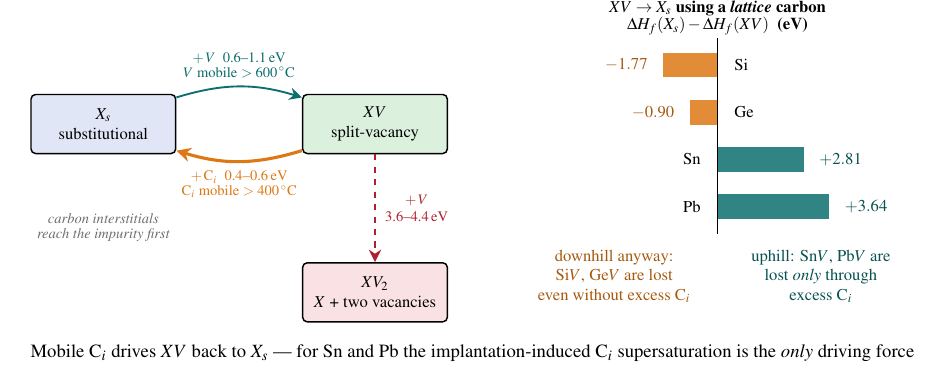}
\end{graphicalabstract}

\begin{highlights}
\item Group-IV--vacancy centers compete with substitutional X$_\text{s}$ and XV$_2$ during annealing
\item Carbon interstitials convert XV back to X$_\text{s}$ with barriers of only 0.36-0.56~eV
\item Carbon interstitial is mobile $\sim200$~$^\circ$C below the vacancy, so annealing erodes the split vacancy
\item For Sn and Pb this loss is driven solely by implantation-induced excess carbon interstitials
\item Coulomb blocking of C$_i$ capture needs E$_\text{F}$ below 0.2--0.9~eV or above 4.3~eV
\end{highlights}

\begin{keyword}
SiV  center \sep GeV center \sep SnV center \sep PbV center  \sep Ion implantation  \sep Density functional theory
\end{keyword}

\end{frontmatter}



\section{Introduction}
\label{sec1}
Quantum emitters play an important role in quantum computing and quantum information processing \cite{awschalom_quantum_2018,atature_material_2018,aharonovich_solid-state_2016,iwasaki_color_2020}. In recent years, color centers in diamond have emerged as promising solid-state quantum emitters \cite{doherty_nitrogen-vacancy_2013,abobeih_one-second_2018,thiering_color_2020}. Among them, the negatively charged nitrogen-vacancy color center (N$V^-$) has been extensively studied \cite{gruber_scanning_1997,wrachtrup_quantum_2001,gali_ab_2019,waldherr_dark_2011,beha_optimum_2012,casola_probing_2018,norman_novel_2021,gaebel_room-temperature_2006,santori_nanophotonics_2010}, because its spin state can be manipulated, initialized, and read out at room temperature. However, the $NV$ center suffers from a small Debye-Waller factor and weak fluorescence emission in the zero-phonon line (\textasciitilde3\%), which has triggered the search for alternative candidates with improved optical properties.

The group-IV--vacancy color centers, denoted by $XV$ centers, have attracted considerable attention \cite{hepp_electronic_2014,siyushev_optical_2017,rugar_quantum_2021,trusheim_lead-related_2019,thiering_ab_2018}. In the split-vacancy configuration, the group-IV impurity occupies an interstitial position between two adjacent carbon vacancies in diamond, giving the 
$XV$ center inversion symmetry. The negatively charged $XV^-$ center has spin 1/2 and exhibits favorable optical properties \cite{chen_building_2020,arjona_martinez_photonic_2022,bradac_quantum_2019}, including a large Debye-Waller factor, strong zero-phonon line emission, a short radiative lifetime, and spectral stability.

These color centers in diamond are typically fabricated by ion implantation or incorporated during diamond growth \cite{iwasaki_color_2020,bradac_quantum_2019,schroder_scalable_2017,ju_preparations_2021,wang_single_2006,neu_single_2011,iwasaki_germanium-vacancy_2015,tchernij_single-photon-emitting_2017,iwasaki_tin-vacancy_2017,ditalia_tchernij_single-photon_2018,trusheim_lead-related_2019}. Among these approaches, ion implantation is widely used because it enables control over the number and spatial distribution of the implanted impurities. However, ion implantation inevitably introduces intrinsic defects and lattice disorder, which can affect the formation yield and optical properties of the color centers.

Experiments on implanted Ge and Sn have shown that implanted impurities can occupy both split-vacancy and substitutional configurations, and that their relative populations evolve during annealing. For implanted Ge, annealing decreases the split-vacancy configuration fraction and increases the substitutional fraction. This redistribution has been suggested to result from the capture of mobile carbon interstitials by Ge$V$ \cite{wahl_structural_2024}. For implanted Sn, annealing also converts part of the Sn population from the split-vacancy configuration to substitutional sites, but this conversion is less pronounced than in implanted Ge, and a substantial fraction of Sn remains in the split-vacancy configuration after annealing \cite{wahl_direct_2020}. An additional emission near 595 nm has been reported in Sn-implanted diamond, although the microscopic structure of this new optical center remains unresolved \cite{cheng_laser_2025,pugliese_photoactivation_2025,iwasaki_tin-vacancy_2017}. A Sn$V$--C$_i$ complex has been proposed as a possible model for this additional optical emission~\cite{cheng_laser_2025}. At the atomic scale, molecular-dynamics simulations of Si implantation have examined implantation-induced damage and vacancy evolution during annealing and found partial recovery of the damaged diamond structure at high temperature~\cite{fu_molecular_2019}. First-principles calculations have shown that implantation-induced vacancies can form complexes with $XV$ centers and substantially modify their properties \cite{kuate_defo_how_2019}. Studies of N$V$--C$_i$ and Sn$V$--C$_i$ systems in diamond have suggested that the diffusion and subsequent interaction of C$_i$ with color centers can influence their optical response \cite{chen_simulating_2025,cheng_laser_2025}. Despite these efforts, the microscopic formation mechanisms of group-IV--vacancy centers remain incompletely understood, and a systematic comparison of vacancy- and interstitial-mediated formation and conversion pathways across the Si--Pb series is still lacking.

In this work, we use first-principles calculations to investigate the roles of vacancies and carbon interstitials in the formation and structural stability of group-IV--vacancy centers in diamond. Formation energies and binding energies quantify the thermodynamic driving forces and relative defect stability, while diffusion calculations distinguish kinetically accessible pathways from thermodynamically favorable but kinetically hindered ones. Electronic structures and charge transition levels are further analyzed. The results reveal a competing reaction network governing the formation and conversion of split-vacancy configurations and indicate how Fermi-level control could suppress undesirable conversions and stabilize the split-vacancy configurations.

\section{Method}
 Density functional theory (DFT) calculations are performed with the Vienna \textit{Ab initio} Simulation Package (VASP) \cite{kresse_efficient_1996} using the projector augmented wave (PAW) method \cite{blochl_projector_1994,kresse_ultrasoft_1999}. To avoid interactions between the defects and their periodic images, all calculations are performed using a 4 $\times$ 4 $\times$ 4 cubic supercell containing 512 atoms and with $\Gamma$-point sampling. The kinetic-energy cutoff for the plane-wave basis is set to 420 eV. The convergence criteria are $10^{-5}$ eV in total energy and 0.01 eV/\AA{} in force. Formation energies, binding energies, ground-state electronic structures of defects, and charge transition levels (CTLs) are calculated using the screened hybrid functional of Heyd-Scuseria-Ernzerhof (HSE06) \cite{heyd_hybrid_2003,heyd_erratum_2006,krukau_influence_2006}. The calculated bandgap (5.4~eV) agrees with the experimental value (5.5~eV) within 0.1 eV. All migration barriers are calculated using the climbing image nudged elastic band (CI-NEB) method \cite{henkelman_climbing_2000} with the Perdew--Burke--Ernzerhof (PBE) functional \cite{perdew_generalized_1996}. For each barrier calculation, NUPDOWN was set based on the final state spin configuration and kept fixed for all images, including both endpoints. All calculations were spin-polarized. During structural relaxation, the lattice vectors and volume are kept fixed. To assess the dependence on the functional, we performed single-point HSE06 calculations at the PBE geometries for the Si$V+V\rightarrow$ Si$V_2$ reaction; the resulting barrier of 4.37~eV differs from the PBE value of 4.23~eV by only 0.14~eV. Moreover, our PBE migration barrier of the isolated $V$ (2.79~eV) agrees with the values of 2.8~eV obtained with HSE06 \cite{deak_formation_2014} and LDA \cite{breuer_ab_1995}.\label{rev:hse}

The defect formation energy ($\Delta$$H_f$) of defect in charge state $q$ is defined as \cite{zhang_chemical_1991,freysoldt_first-principles_2014}:
\begin{equation}
\Delta{H_f}(\alpha,q) = \Delta{E(\alpha,q)} + \sum_{i}n_i\mu_i + qE_\mathrm{F} + E_{\mathrm{corr}}(q)
\end{equation}
$\Delta{E(\alpha,q)} = E(\alpha,q) - E(\mathrm{host}) + \sum_{i}n_iE_i + qE_{\mathrm{VBM}}$. $E(\alpha, q)$ and $E$(host) are the total energy of the supercell with and without defect, respectively. $\alpha$ denotes defect, $q$ indicates charge state, $n_i$ and $\mu_i$ are the number and chemical potential of each constituent $i$. $E_\mathrm{F}$ is the Fermi level referring to the valence band maximum ($E_{\mathrm{VBM}}$) of the host. The correction term $E_{\mathrm{corr}}$ originates from the electrostatic interaction between charged defects in the periodic supercell and is evaluated using the Freysoldt-Neugebauer-Van de Walle (FNV) method \cite{freysoldt_fully_2009}.

The thermodynamic charge transition level $\varepsilon(q/q^{'})$ is the Fermi level position at which the formation energies of the charge states, $q$ and $q^{'}$, are equal.
\begin{equation}
\varepsilon(q/q^{'}) = \frac{E(\alpha,q) - E(\alpha,q^{'}) + E_{\mathrm{corr}}(q)-E_{\mathrm{corr}}(q^{'})}{(q^{'}-q)} - E_{\mathrm{VBM}}
\end{equation}

Binding energies quantify the thermodynamic driving force for defect reactions and are independent of the atomic chemical potentials. Considering the mobility of C$_i$ and $V$ in diamond during annealing \cite{iakoubovskii_annealing_2003,allers_annealing_1998}, we focused on three main reactions and defined their binding energies as:
\begin{align}
E^{b1} &= \Delta{H_f}(X_s) + \Delta{H_f}(V) - \Delta{H_f}(XV)\\
E^{b2} &= \Delta{H_f}(XV) + \Delta{H_f}(\mathrm{C}_i) - \Delta{H_f}(X_s)\\
E^{b3} &= \Delta{H_f}(XV) + \Delta{H_f}(V) - \Delta{H_f}(XV_2)
\end{align}

In general, defects may exist in different charge states depending on the Fermi level, which can introduce extra electrons or holes during reactions. In this study, only neutral charge states are considered to clarify the trend; the influence of the Fermi level on these reactions is treated separately through the charge transition levels in Sec.~\ref{subsec3}. Binding energies are evaluated using the formation energies of the respective defects in their neutral ground states, with a broken-symmetry singlet solution for the neutral vacancy. With the definitions above, a positive binding energy indicates that the corresponding reaction is exothermic and energetically favorable. 

\section{Results and discussion}
\subsection{Formation and binding energies}
\label{subsec1}
The calculated formation energies are summarized in Table~\ref{table1}. Sn$V$ and Pb$V$ have lower formation energies than their substitutional defects ($X_s$), whereas Si$V$ and Ge$V$ have higher formation energies than Si$_s$ and Ge$_s$. This can be attributed to atomic size effects \cite{goss_vacancy-impurity_2005}, as larger $X$ impurities induce stronger lattice compression. Specifically, the first nearest neighbor C--C bond lengths are reduced by 1.43\% in Si$_s$, 1.81\% in Ge$_s$, 2.38\% in Sn$_s$, and 2.51\% in Pb$_s$, respectively, relative to the perfect lattice. Shorter C--C bonds enhance bond strength, increasing the energy cost to create a vacancy. When the released strain is insufficient to compensate for this cost, the formation energy of $X_s$ exceeds that of $XV$. We note that the carbon chemical potential is fixed by the host diamond crystal ($\mu_\mathrm{C}=0$ corresponds to bulk diamond), while the impurity chemical potentials are referenced to the elemental solids ($\mu_X=0$). Since no thermodynamically stable bulk Ge, Sn, or Pb carbide exists, the elemental reservoir is the natural and essentially unique reference for these species, and carbide precipitation does not compete with $XV$ formation. For Si, by contrast, the stable SiC phase bounds the accessible range of $\mu_\mathrm{Si}$ under carbon-rich conditions and provides a competing precipitation channel. The absolute formation energies and the inferred concentrations still depend on the impurity chemical potentials realized during growth or implantation, so the cross-impurity comparison is best read as a relative trend. Formation energies of this magnitude imply negligible equilibrium concentrations: even for Si$V$, which has the lowest formation energy among the $XV$ centers (6.92~eV under C-rich conditions without doping), the equilibrium concentration at 900~$^\circ$C remains extremely low. The $XV$ centers are therefore products of non-equilibrium processes, such as ion implantation, which supplies the impurities together with vacancies and interstitials, or incorporation at the growth surface during chemical vapor deposition. Their formation is governed by the binding energies and migration barriers of the constituents rather than by the absolute formation energies.\label{rev:eq}

\begin{table}[htbp]
\centering
\caption{Formation energies of $X_s$, $XV$, $XV_2$, $V$, and C$_i$ in the neutral charge states. For Si$_s$, Si$V$ and Si$V_2$, the formation energies are calculated under C-rich conditions. For the other defects, the chemical potentials $\mu{\mathrm{_C}}$ and $\mu_X$ are set to 0, as explained in the main text.}\label{table1}
\begin{tabular}{c c c c}
  \hline\hline
  Defect & $\Delta {H_f}$ (eV) & $q$ & $\mu$ (eV)\\
  \hline
  Si$_s$& 5.15 & 0 & C-rich \\
  Ge$_s$& 7.16 & 0 & $\mu_{\mathrm{C},\,\mathrm{Ge}}=0$\\
  Sn$_s$& 14.12 & 0 & $\mu_{\mathrm{C},\,\mathrm{Sn}}=0$\\
  Pb$_s$& 17.84& 0 & $\mu_{\mathrm{C},\,\mathrm{Pb}}=0$\\
  \hline
  Si$V$ & 6.92& 0 & C-rich \\
  Ge$V$ & 8.06& 0 & $\mu_{\mathrm{C},\,\mathrm{Ge}}=0$\\
  Sn$V$ & 11.31& 0 & $\mu_{\mathrm{C},\,\mathrm{Sn}}=0$ \\ 
  Pb$V$ & 14.20 & 0 & $\mu_{\mathrm{C},\,\mathrm{Pb}}=0$\\
  \hline
  Si$V_2$ & 10.18 & 0 & C-rich\\
  Ge$V_2$ & 11.43 & 0 & $\mu_{\mathrm{C},\,\mathrm{Ge}}=0$\\
  Sn$V_2$ & 14.12 & 0 & $\mu_{\mathrm{C},\,\mathrm{Sn}}=0$\\
  Pb$V_2$ & 16.42& 0 & $\mu_{\mathrm{C},\,\mathrm{Pb}}=0$\\
  \hline
  C$_i$ & 11.75& 0 & $\mu_\mathrm{C}=0$ \\
  \hline
  $V$ & 6.93& 0 & $\mu_\mathrm{C}=0$ \\
  \hline\hline 
\end{tabular}
\end{table}

\begin{table}[htbp]
\centering 
\caption{Binding energies ($E^b$) for the defect reactions $X_s + V \rightarrow XV$ ($E^{b1}$), $XV + \mathrm{C}_i \rightarrow X_s$ ($E^{b2}$), and $XV + V \rightarrow XV_2$ ($E^{b3}$).}\label{table2}
\begin{tabular}{c c c c c}
  \hline\hline
  $E^b$ (eV) &Si &Ge &Sn &Pb \\ 
  \hline
  $E^{b1}$ &5.17 &6.03 &9.75 &10.58 \\
  $E^{b2}$ &13.51 &12.65 &8.94 &8.10 \\
  $E^{b3}$  &3.67 &3.57 &4.12 &4.72 \\
  \hline\hline 
\end{tabular}
\end{table}

\begin{figure*}
\centering
\includegraphics[width=\linewidth]{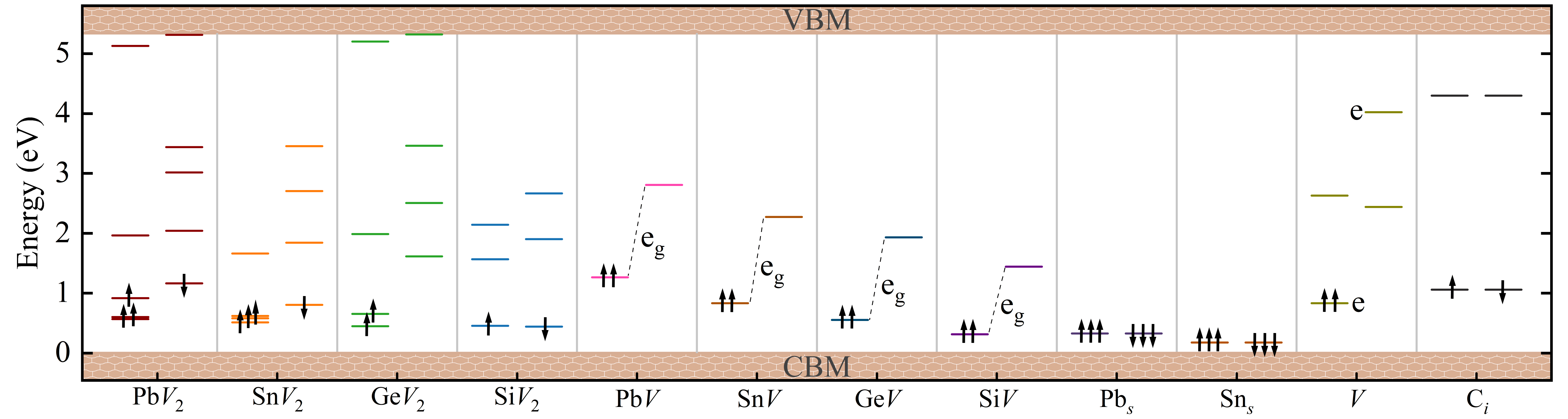} 
\caption{HSE06 Kohn-Sham levels of neutral defects in their ground states at the $\Gamma$ point in diamond. Only states within the fundamental bandgap are shown. Si$_s$ and Ge$_s$ do not introduce the localized in-gap states and are therefore absent. Occupied majority- and minority- spin states are denoted by $\uparrow$ and $\downarrow$, respectively. }\label{Fig1}
\end{figure*}

Despite relatively high formation energies, the large positive binding energies indicate that the formation of $XV$ is energetically favorable, as listed in Table~\ref{table2}. In addition, inevitable vacancies and self-interstitials introduced by ion implantation can further interact with $XV$. The positive binding energies indicate that the conversion of $XV$ to $X_s$, as well as the formation of the complex defect $XV_2$, is thermodynamically favorable. Moreover, for the reaction $X_s + V \rightarrow XV$, the binding energies for Si and Ge are smaller than the formation energy of $V$. Similarly, for the reaction $XV + \mathrm{C}_i \rightarrow X_s$, the binding energies for Sn and Pb are smaller than the formation energy of C$_i$. These two statements are equivalent to the sign of $\Delta H_f(XV)-\Delta H_f(X_s)=E^{b2}-\Delta H_f(\mathrm{C}_i)$, which is $+1.77$, $+0.90$, $-2.81$ and $-3.64$ eV for Si, Ge, Sn and Pb, respectively. Both conversions are exothermic for every impurity, so neither configuration is thermodynamically excluded and the populations are determined kinetically rather than by an equilibrium between them. The sign above instead identifies where the driving force comes from. Because C$_i$ is itself a high-energy species ($\Delta H_f = 11.75$ eV), the conversion $XV \rightarrow X_s$ is downhill for all four impurities; if the carbon were taken from the surrounding lattice instead of from an interstitial, the same conversion would remain downhill by 1.77 and 0.90 eV for Si and Ge, but would cost 2.81 and 3.64 eV for Sn and Pb. The loss of Sn$V$ and Pb$V$ is therefore driven entirely by the interstitial supersaturation created during implantation, whereas Si$V$ and Ge$V$ are unstable against substitutional incorporation even in its absence. Suppressing the interstitial supersaturation should accordingly close this loss channel for Sn and Pb, but not for Si and Ge. Since binding energies are derived from formation energies, the variations can be consistently explained by atomic size effects. These trends indicate that impurity size affects both defect solubility and the energetic driving forces of defect reactions. 

\begin{figure}[htbp]
\centering
\includegraphics[width=\linewidth]{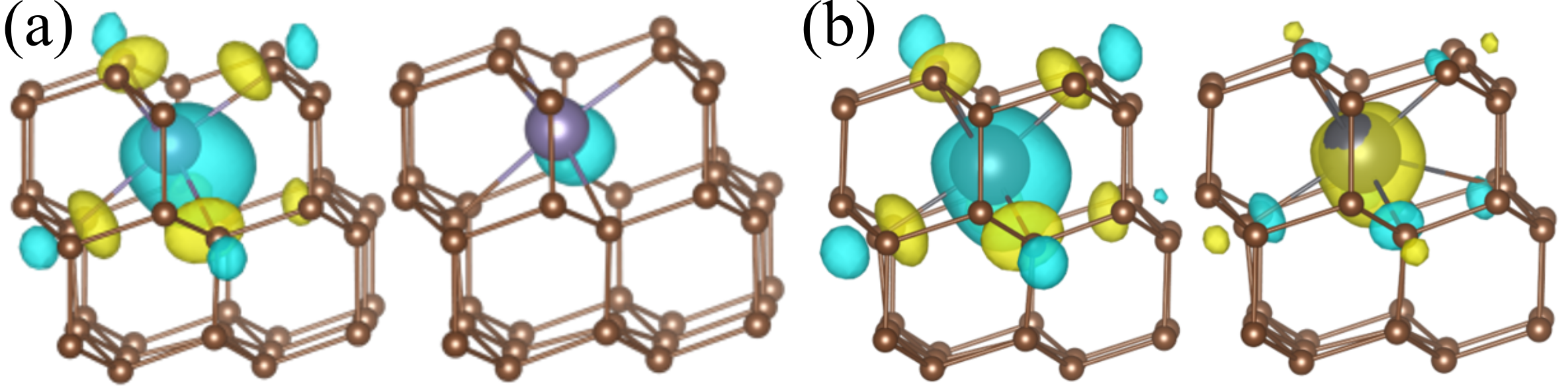} 
\caption{Defect state wave functions of neutral (a) Ge$V_2$ and (b) Pb$V_2$ viewed along the [111] direction. Only the in-gap states close to the CBM are shown; the corresponding states of Si$V_2$ and Sn$V_2$ lie above the CBM and are therefore omitted. In each panel, the left and right isosurfaces correspond to the up-spin and down-spin components, respectively. The isosurface value is 4 $\times$ 10$^{-7}$~e/\AA{}$^3$.}\label{Fig2}
\end{figure}

Furthermore, we analyze the ground state electronic structures of these defects and the in-gap Kohn-Sham levels are shown in Fig.~\ref{Fig1}. For $X_s$ defects with $T_{d}$ symmetry, no localized defect states are introduced within the bandgap for Si$_s$ and Ge$_s$. In contrast, Sn$_s$ and Pb$_s$ exhibit occupied, triply degenerate $t_2$ Kohn-Sham levels within the bandgap, at $E_{\mathrm{VBM}}+0.18$ eV and $E_{\mathrm{VBM}}+0.33$ eV, respectively. The upward shift of the defect state can be attributed to the weakening $p$--$p$ orbital coupling between the $X$ atom and its neighboring carbon atoms with increasing atomic number of the $X$ atom. As a result, the occupied bonding state is  pushed upwards into the bandgap. However, the $t_2$ level of Sn$_s$ is not a localized defect state but rather a resonant state hybridized with the valence band. 

The $XV$ centers with $D_{3d}$ symmetry exhibit triplet ground states and doubly degenerate e$_g$ levels in the bandgap. For $XV_2$ with $C_{1h}$ symmetry, more defect states are introduced owing to the presence of additional carbon dangling bonds, compared with $XV$. Moreover, Ge$V_2$ and Pb$V_2$ exhibit an additional defect state close to the CBM, while the corresponding state in Si$V_2$ and Sn$V_2$ lies above the CBM. This difference can be attributed to the lower symmetry of $XV_2$, which allows $s$--$p$ orbital coupling between the $X$ atom and the carbon vacancy. The stronger coupling in Si$V_2$ and Sn$V_2$ shifts this state upward into the conduction band, while the weaker interaction in Ge$V_2$ and Pb$V_2$ leaves it near the CBM within the bandgap. The wave functions of this defect state for Ge$V_2$ and Pb$V_2$ are shown in Fig.~\ref{Fig2}(a) and (b), respectively. Specifically, Ge$V_2$, Sn$V_2$, and Pb$V_2$ exhibit high-spin ground states ($S=1$), whereas Si$V_2$ has a singlet ground state ($S=0$). 

For the isolated $V$, the ground state has been identified as a correlated singlet \cite{deak_formation_2014}. Since this many-electron state cannot be represented by a unique set of Kohn--Sham single-particle levels, Fig.~\ref{Fig1} shows, for clarity, the single-particle levels of a triplet state with $D_{2d}$ symmetry. Following the approach of Ref. \cite{deak_formation_2014}, we also obtained a broken-symmetry singlet solution that is 0.18 eV lower in energy than the triplet solution. The formation energy discussed above refers only to neutral $V$ and is therefore evaluated using the energy of this lower-energy broken-symmetry singlet. For the charge transition levels, however, we follow the same treatment as Ref. \cite{deak_formation_2014}, in which the triplet configuration is used as the reference state for $V^{0}$. In addition, C$_i$ with $D_{2d}$ symmetry has a singlet ground state. 

\subsection{The diffusion behavior}
\label{subsec2}
\begin{figure}[htbp]
\centering
\includegraphics[width=\linewidth]{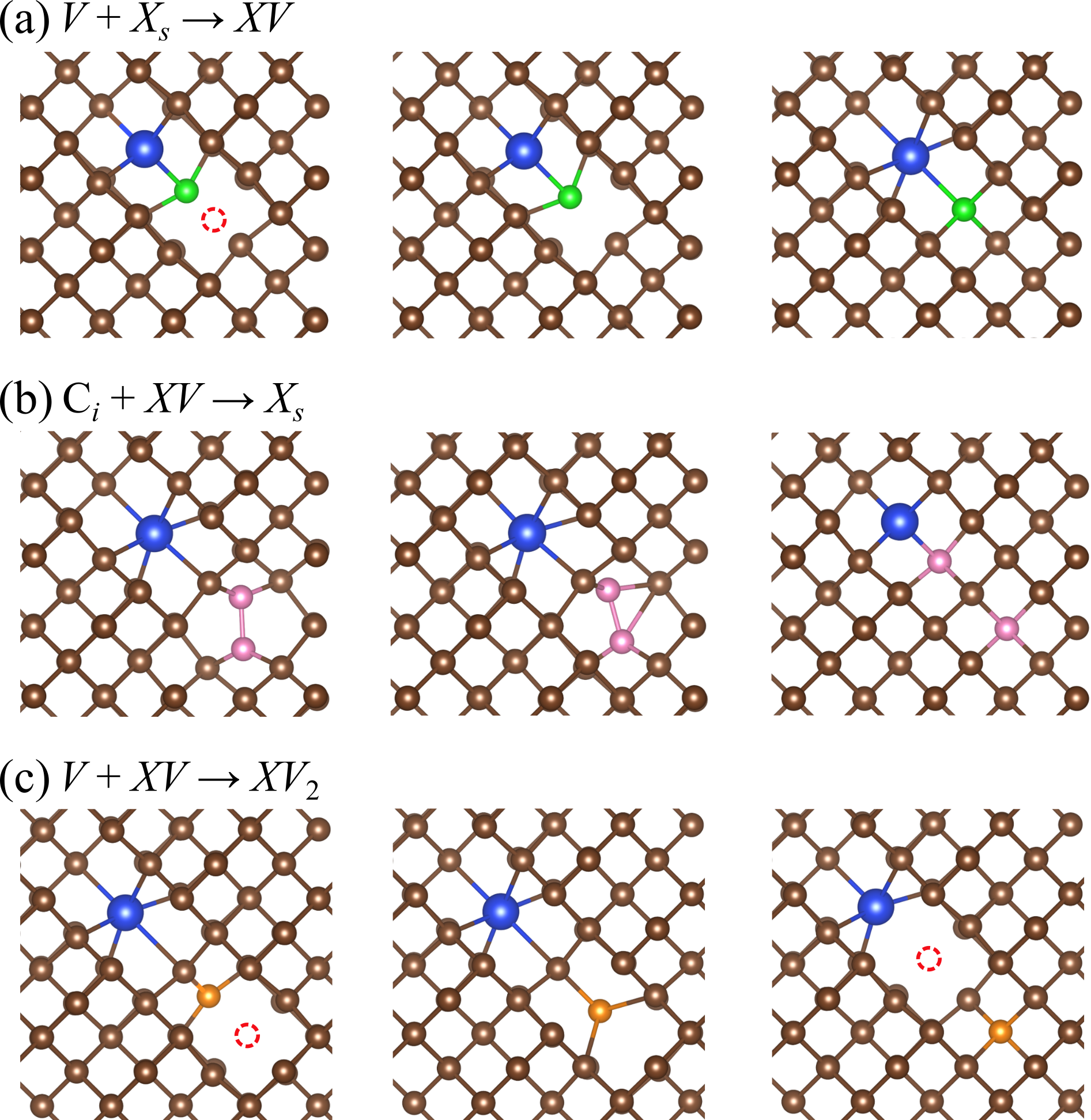} 
\caption{Diffusion pathways for (a) capture of $V$ by $X$ ($V$ is at the second-nearest-neighbor); (b) capture of C$_i$ by $XV$ (C$_i$ is at the second-nearest-neighbor and highlighted in pink); (c) capture of $V$ by $XV$ ($V$ is at the third-nearest-neighbor). The blue sphere and red dashed circle are $X$ atom and $V$, respectively. The C atoms involved in the diffusion process are highlighted by green and orange spheres. The left, middle, and right snapshots correspond to the initial, transition, and final configurations, respectively. The calculated diffusion barriers for isolated C$_i$ and $V$ are 1.89 eV and 2.79 eV, respectively, in agreement with previous DFT studies \cite{deak_formation_2014,chen_simulating_2025,goss_self-interstitial_2001,breuer_ab_1995}.}\label{Fig3}
\end{figure}

\begin{figure}[htbp]
\centering
\includegraphics[width=\linewidth]{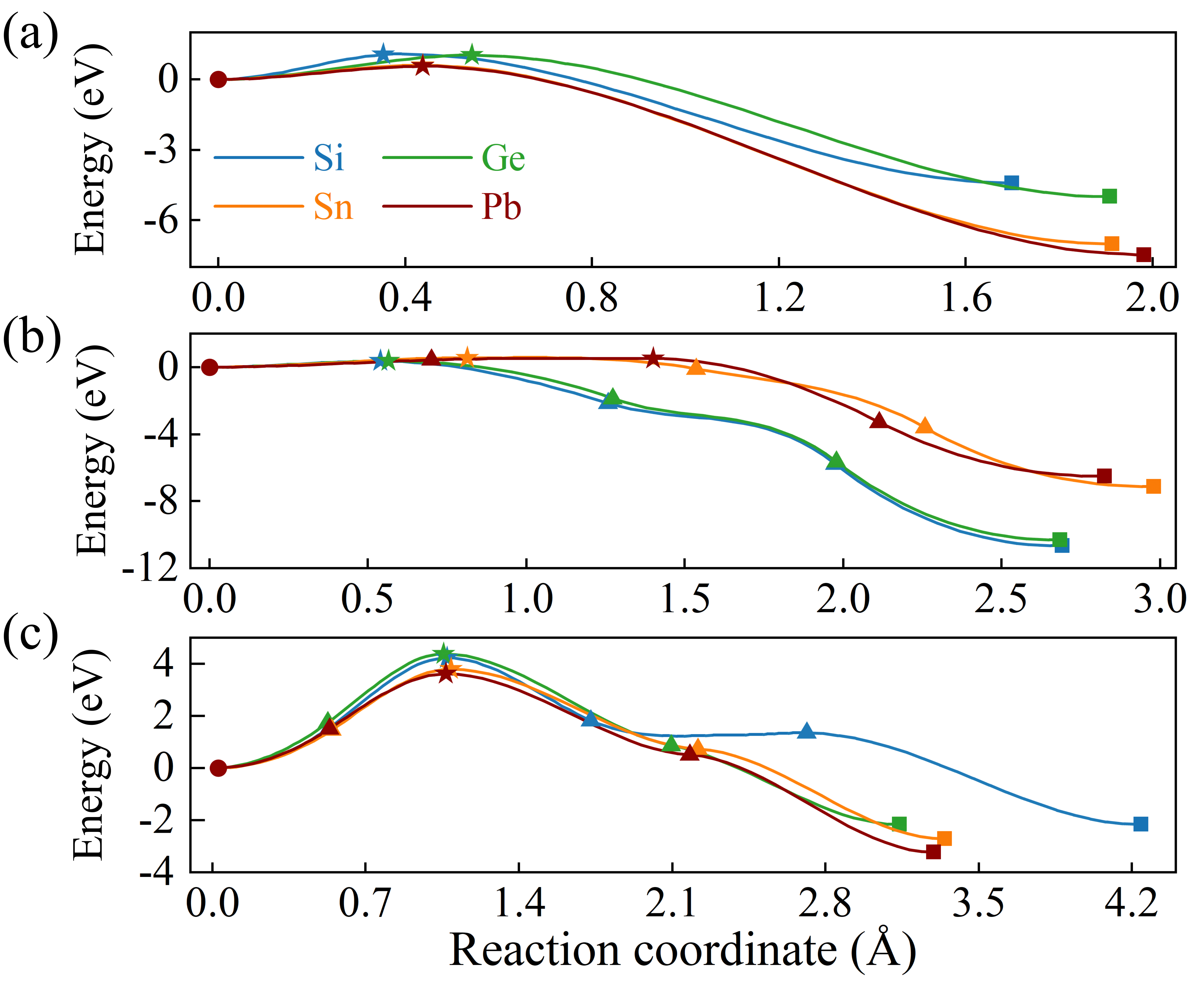} 
\caption{Energy profiles for the three reactions: (a) $X_s + V \rightarrow XV$, (b) $XV + \mathrm{C}_i \rightarrow X_s$, and (c) $XV + V \rightarrow XV_2$. Filled circles ($\bullet$), stars ($\star$), and squares (\scalebox{0.55}{$\blacksquare$}) denote the initial, transition, and final states. The corresponding atomic configurations are shown in Fig.~\ref{Fig3}. Triangles (\scalebox{0.8}{$\blacktriangle$}) denote the remaining intermediate images used in the calculations. One intermediate image was used in (a), and three were used in (b). In (c), four intermediate images were used for $X=\mathrm{Si}$, and three for $X=\mathrm{Ge}$, $\mathrm{Sn}$, and $\mathrm{Pb}$. Energies are given relative to the initial state.}\label{barrier}
\end{figure}

For the formation of $XV$, we adopt a simple model, analogous to that used for nitrogen-vacancy color center in diamond \cite{deak_formation_2014,mainwood_nitrogen_1994,naydenov_increasing_2010,pezzagna_creation_2010,rabeau_implantation_2006}, where an $X$ ion captures a mobile $V$ during ion implantation. Representative initial, saddle-point, and final configurations along this pathway are shown in Fig.~\ref{Fig3}(a), and the corresponding energy profiles are presented in Fig.~\ref{barrier}(a). The calculated barriers are summarized in Table~\ref{tab:barriers}. Our calculations show that the reaction barrier is quite low, indicating that $XV$ forms readily. Despite this, the experimentally observed formation yield remains relatively low \cite{baba_annealing_2025,tzeng_improving_2024}, suggesting that interactions with other defects may influence the process. Experiments show that both $V$ and C$_i$ can migrate in diamond. C$_i$ has lower migration barriers and activation temperatures than $V$ \cite{iakoubovskii_annealing_2003,allers_annealing_1998}. This indicates that both defects can interact with $XV$ and affect its formation. Previous work has mainly focused on $V$ diffusion in the presence of $XV$ \cite{kuate_defo_how_2019}. To clarify the contributions of both defects, we systematically investigated the diffusion of C$_i$ and $V$ in the presence of $XV$.

For the capture of C$_i$ by $XV$, the lowest-energy split-<001> C$_i$ configuration migrates towards the $XV$ and recombines to form $X_s$. Representative configurations along the recombination pathway and the corresponding energy profiles are shown in Fig.~\ref{Fig3}(b) and Fig.~\ref{barrier}(b), respectively, with the calculated barriers summarized in Table~\ref{tab:barriers}. The diffusion barrier is lower than that of the isolated C$_i$, indicating that recombination with $XV$ is favorable along this pathway. The reported annealing-induced redistribution is consistent with this behavior \cite{wahl_structural_2024,wahl_direct_2020}: for Ge implantation, \textasciitilde20\% of ions adopt the split-vacancy configuration and \textasciitilde45\% are substitutional at room temperature; after annealing to 900 $^\circ\mathrm{C}$, the split-vacancy fraction decreases to 6--9\%, while the substitutional fraction rises to 85--96\%. For Sn implantation,  approximately 40\% of ions initially adopt the split-vacancy configuration, while approximately 30\% remain in the ideal split-vacancy configuration after annealing at 920 $^\circ\mathrm{C}$. These observations are consistent with the calculated small recombination barriers. Additionally, the experimental observation of the type-II Sn defect has been hypothesized to be a complex defect formed by Sn$V$ and C$_i$ \cite{cheng_laser_2025}. Theoretical work has also linked C$_i$ dynamics to photoluminescence blinking in N$V$ centers \cite{chen_simulating_2025}. These results highlight the significant influence of C$_i$ on the structural stability and on the resulting density of optically active $XV$ centers. The conversion of $XV$ into $X_s$ competes directly with $XV$ formation. We note, however, that all of these recombination barriers lie 1.3--1.5 eV below the migration barrier of the isolated C$_i$, so that at 1173 K the capture step is faster than a further migration hop by five to seven orders of magnitude. The conversion is thus diffusion-limited for all four impurities: essentially every C$_i$ that reaches an $XV$ center is captured, and the competition between the two channels is governed by which mobile species arrives first rather than by the local barriers. 

\begin{table}[htbp]
\centering 
\caption{Calculated migration/recombination barriers (in eV) for the three defect reactions, obtained with the CI-NEB method at the PBE level. The rate-limiting step of $V+XV\rightarrow XV_2$ corresponds to $V$ at the third-nearest-neighbor site. For reference, the isolated migration barriers are 1.89 eV (C$_i$) and 2.79 eV ($V$).}\label{tab:barriers}
\begin{tabular}{c c c c c}
  \hline\hline
  Reaction &Si &Ge &Sn &Pb \\
  \hline
  $X_s+V\rightarrow XV$ &1.06 &1.03 &0.60 &0.56 \\
  $\mathrm{C}_i+XV\rightarrow X_s$ &0.36 &0.37 &0.56 &0.52 \\
  $V+XV\rightarrow XV_2$ &4.23 &4.39 &3.80 &3.62 \\
  \hline\hline
\end{tabular}
\end{table}

For the capture of $V$ by $XV$, when $V$ is at the second-nearest-neighbor of $XV$ (the corresponding C atom highlighted in orange), the configuration is unstable and spontaneously relaxes to $XV_2$ for Ge, Sn, and Pb. For Si, a negligible diffusion barrier of 0.08 eV enables spontaneous conversion under thermal fluctuations or perturbation. However, formation of $XV_2$ is limited by an extremely high barrier (\textasciitilde4 eV) when $V$ is at the third-nearest-neighbor. Representative configurations and the corresponding energy profiles are shown in Fig.~\ref{Fig3}(c) and Fig.~\ref{barrier}(c), respectively, and the calculated barriers are summarized in Table~\ref{tab:barriers}. Consequently, the conversion of $XV$ into $X_s$ mediated by C$_i$ is expected to be dominant. The electronic structure of Si$V_2$ has been studied theoretically \cite{goss_density_2007,thiering_complexes_2015}, and experiment has identified the WAR3 EPR center as a complex defect formed by Si$V_2$ bound to hydrogen (Si$V_2$:H) \cite{dhaenens-johansson_epr_2010}. To our knowledge, no other $XV_2$ defects have been experimentally identified. This may result from the high vacancy-capture barrier or optical emission in the near-infrared region that has not been detected. Previous theoretical work also suggested that Ge$V^-$ exhibits a higher low-frequency phonon density than Si$V^-$, which would explain its shorter spin coherence time despite the larger ground-state splitting. This work proposed that introducing vacancies could reduce the acoustic phonon density and potentially enhance spin coherence times \cite{kuate_defo_how_2019}.

To provide a quantitative estimate of timescales associated with the calculated barriers, we convert these barriers into rates using the Arrhenius expression $k=\nu\exp[-E_a/k_{\mathrm{B}}T]$, with an assumed prefactor of $\nu=10^{13}$ $\mathrm{s}^{-1}$ at $T=1173$ K ($900\,^{\circ}$C). For $X_s+V \rightarrow XV$, the estimated rates are approximately $2.8\times10^8$ $\mathrm{s}^{-1}$, $3.8\times10^8$ $\mathrm{s}^{-1}$, $2.6\times10^{10}$ $\mathrm{s}^{-1}$, and $3.9\times10^{10}$ $\mathrm{s}^{-1}$ for Si, Ge, Sn, and Pb, respectively. The corresponding rates for $XV+\mathrm{C}_i \rightarrow X_s$ are $2.8\times10^{11}$ $\mathrm{s}^{-1}$, $2.6\times10^{11}$ $\mathrm{s}^{-1}$, $3.9\times10^{10}$ $\mathrm{s}^{-1}$, and $5.8\times10^{10}$ $\mathrm{s}^{-1}$, respectively. In contrast, the estimated rates for $XV+V \rightarrow XV_2$ are substantially lower, at $6.7\times10^{-6}$ $\mathrm{s}^{-1}$, $1.4\times10^{-6}$ $\mathrm{s}^{-1}$,$4.7\times10^{-4}$ $\mathrm{s}^{-1}$, and $2.8\times10^{-3}$ $\mathrm{s}^{-1}$, respectively. These  results indicate that $XV$ formation and its conversion to $X_s$ are kinetically accessible, while the conversion of $XV$ to $XV_2$ is comparatively less accessible at the experimental annealing temperature.

Spin conservation does not alter these conclusions. With the ground-state spins of Sec.~\ref{subsec1}, the capture of $V$ by Ge$V$, Sn$V$, and Pb$V$ (an $S=1$ and an $S=0$ reactant forming an $S=1$ product) is spin-allowed, whereas $X_s+V\rightarrow XV$, $XV+\mathrm{C}_i\rightarrow X_s$, and Si$V+V\rightarrow$ Si$V_2$ connect initial and final states of different total spin. At 900~$^\circ$C, however, the triplet state of $V$, which lies only 0.18~eV above the broken-symmetry singlet, is populated to about one third, and the $S=1$ R2 EPR center has been assigned to the neutral split interstitial \cite{goss_self-interstitial_2001}, showing that a triplet state of C$_i$ is also accessible. Spin-allowed channels are therefore thermally available for all three reactions. A residual spin restriction can reduce the effective attempt frequency, but it does not change the barriers or their ordering. In particular, the capture of C$_i$ is faster than a further migration hop by five to seven orders of magnitude, so even a reduction by two to three orders of magnitude would leave the conversion diffusion-limited.\label{rev:spin}

The 0.90 eV difference between the migration barriers of C$_i$ and $V$ also has a direct consequence for annealing protocols. Between approximately 460 and 810 $^\circ$C for the present barriers, and between approximately 350 and 650 $^\circ$C using the experimental migration energies of 1.68 eV for C$_i$ \cite{allers_annealing_1998} and 2.3$\pm$0.3 eV for $V$ \cite{davies_vacancy-related_1992}, C$_i$ is mobile while $V$ is not. In this window the conversion of $XV$ into $X_s$ proceeds without the compensating capture of $V$ by $X_s$, so that time spent there is expected to reduce the $XV$ yield irreversibly. The two emission-channeling experiments discussed above differ in exactly this respect: the Sn sample was annealed in a single step at 920 $^\circ$C \cite{wahl_direct_2020}, whereas the Ge sample was held at 300 and 600 $^\circ$C before reaching 900 $^\circ$C \cite{wahl_structural_2024}. A single-step anneal of Ge-implanted diamond would therefore test whether the larger loss of Ge$V$ is intrinsic to the impurity or a consequence of the annealing schedule.

\begin{figure}[htbp]
\centering
\includegraphics[width=\linewidth]{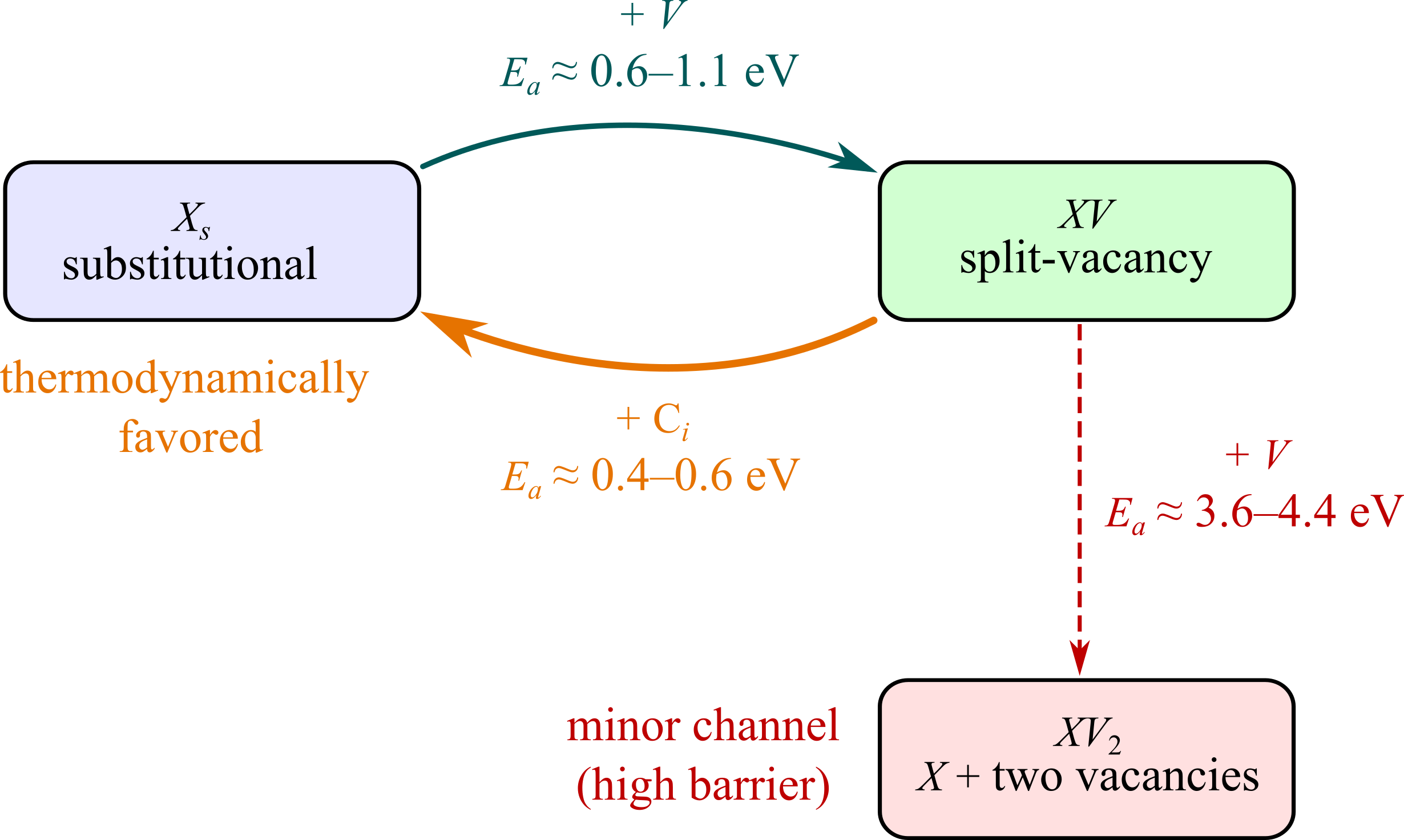} 
\caption{Competing defect-reaction network controlling the formation and stability of $XV$ centers in diamond. A substitutional impurity $X_s$ captures a mobile vacancy to form the split-vacancy $XV$ (teal), while $XV$ is converted back to $X_s$ by capturing the more mobile carbon interstitial C$_i$ (orange); The latter conversion is thermodynamically favored (Table~\ref{table2}), so $X_s$ acts as the dominant sink. Capture of a second vacancy to form $XV_2$ (red, dashed) is gated by a high barrier and remains a minor channel. The activation-energy ranges ($E_a$) span the Si--Pb series and are calculated using CI-NEB at the PBE level (Table~\ref{tab:barriers}).}\label{fig:network}
\end{figure}

Overall, the three reactions define a competing network, $X_s \rightleftharpoons XV \rightarrow XV_2$ (Fig.~\ref{fig:network}), in which the mobile species C$_i$ and $V$ mediate the conversions. The migration barriers are collected in Table~\ref{tab:barriers}. Forward formation of $XV$ from a substitutional $X_s$ and a mobile $V$ is facile (0.56--1.06 eV), but the reverse conversion driven by the more mobile C$_i$ is comparably easy (0.36--0.56 eV) and thermodynamically strongly favored (Table~\ref{table2}), whereas the $XV_2$ channel is gated by a high barrier ($\sim$3.6--4.4 eV). This naturally reconciles the seemingly opposite effects of annealing: mobile vacancies build up $XV$, while abundant and more mobile C$_i$ released near the end of the implantation range drive $XV$ back to $X_s$. The net outcome is that $X_s$ formation tends to dominate, consistent with the experimentally observed decline of the split-vacancy fraction upon annealing \cite{wahl_structural_2024,wahl_direct_2020,baba_annealing_2025} and with the generally low yield of $XV$ centers.

\subsection{Fermi-level dependence of $XV$ conversion}
\label{subsec3}
Based on the analysis above, C$_i$ and $V$ can influence the structural stability of $XV$ centers. Because each reactant can exist in several charge states, the Fermi level determines which charge states meet in a given reaction, and therefore whether the neutral-state results of Secs.~\ref{subsec1} and \ref{subsec2} apply.

In general, three cases can be distinguished for a reaction between two point defects. (i) If both reactants carry charges of the same sign, their Coulomb repulsion opposes the capture. For two singly charged defects, the screened repulsion $e^2/(4\pi\varepsilon_0\varepsilon_r r)$ with $\varepsilon_r=5.7$ amounts to about 1.0~eV at the second-neighbor distance (2.52~\AA{}) and still to about 0.5~eV at 5~\AA{}, which reduces the capture rate by two to four orders of magnitude at 900~$^\circ$C; such reactions are effectively blocked. (ii) If the reactants carry charges of opposite sign, the Coulomb attraction enhances the capture. (iii) If at least one reactant is neutral, there is no long-range Coulomb interaction, and the approach of the mobile species is governed by its ordinary migration. When both reactants are neutral, the binding energies of Sec.~\ref{subsec1} and the barriers of Sec.~\ref{subsec2} apply directly. When only one of them is charged, the extra charge localized on that defect may still modify the short-range barrier, which we have not calculated.

\begin{figure}[htbp]
\centering
\includegraphics[width=\linewidth]{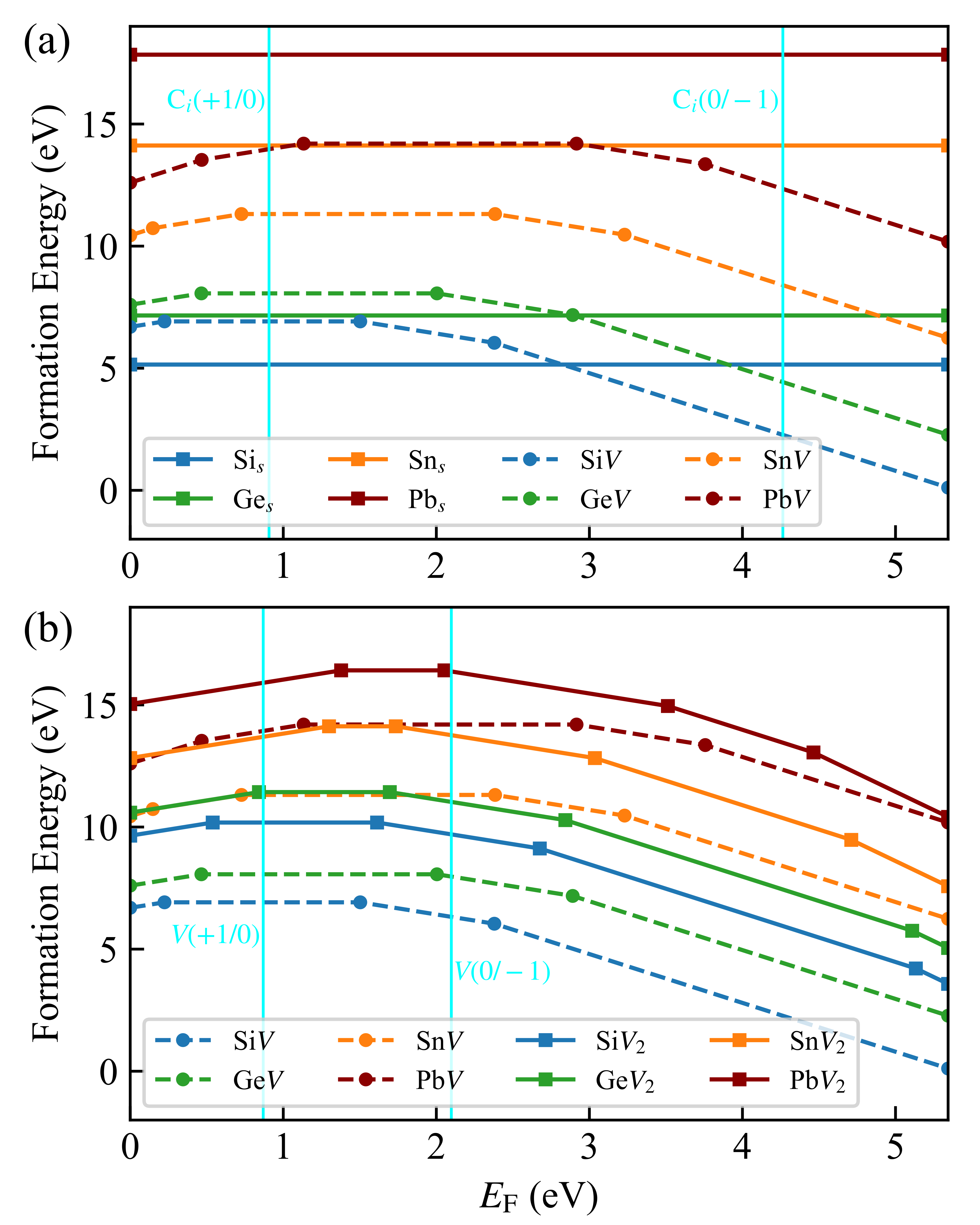} 
\caption{(a) and (b) formation energies as a function of $E_\mathrm{F}$. The zero of the Fermi level is set at the VBM. The formation energies of Si$_s$, Si$V$, and Si$V_2$ are calculated under C-rich conditions, and those of the other defects are calculated with $\mu_{\mathrm{C}}=0$ and $\mu_X=0$. Pb$_s$ has a $\varepsilon$(+1/0) level close to the VBM ($E_\mathrm{VBM}+0.15$ eV), but it lies outside the $E_\mathrm{F}$ range of interest and is therefore not shown. The calculated CTLs of $V$ and $XV$ are in agreement with previous studies \cite{deak_formation_2014,thiering_ab_2018}.}\label{Fig5}
\end{figure}

For the present reactions, case (ii) never occurs. All donor levels of $XV$ [$\varepsilon(+1/0)\leq 1.13$~eV] lie below the acceptor levels of $V$ [$\varepsilon(0/-1)=2.10$~eV] and C$_i$ [$\varepsilon(0/-1)=4.26$~eV], and all acceptor levels of $XV$ [$\varepsilon(0/-1)\geq 1.50$~eV] lie above the donor levels of $V$ [$\varepsilon(+1/0)=0.87$~eV] and C$_i$ [$\varepsilon(+1/0)=0.91$~eV] (Fig.~\ref{Fig5}). Hence, whenever $XV$ is positively charged, its partner is positive or neutral, and whenever $XV$ is negatively charged, its partner is negative or neutral. For each reaction, the $E_\mathrm{F}$ axis therefore divides into a window in which both reactants are neutral, flanking ranges in which only one of them is charged, and, closer to the band edges, ranges in which both carry charges of the same sign and the reaction is Coulomb-blocked. These ranges are summarized in Table~\ref{tab:efwindows}.

For the capture of C$_i$ by $XV$, both reactants are neutral between the $\varepsilon(+1/0)$ of C$_i$ (0.91~eV; for Pb, the $\varepsilon(+1/0)$ of Pb$V$ at 1.13~eV) and the $\varepsilon(0/-1)$ of $XV$ (1.50, 2.00, 2.38, and 2.92~eV for Si, Ge, Sn, and Pb, respectively), as shown in Fig.~\ref{Fig5}(a). The reaction is Coulomb-blocked only when both reactants are positively charged, i.e., for $E_\mathrm{F}$ below the $\varepsilon(+1/0)$ of $XV$ (0.22, 0.47, and 0.73~eV for Si, Ge, and Sn) or of C$_i$ (0.91~eV for Pb), or when both are negatively charged, i.e., for $E_\mathrm{F}$ above the $\varepsilon(0/-1)$ of C$_i$ at 4.26~eV. In between, a neutral C$_i$ meets a charged $XV$, or a positively charged C$_i$ meets a neutral $XV$, and the capture is not suppressed by Coulomb repulsion.

For the capture of $V$ by $XV$, both reactants are neutral between 0.87~eV [the $\varepsilon(+1/0)$ of $V$; 1.13~eV for Pb] and 1.50, 2.00, 2.10, and 2.10~eV for Si, Ge, Sn, and Pb, respectively [Fig.~\ref{Fig5}(b)]. This reaction is Coulomb-blocked below 0.22, 0.47, 0.73, and 0.87~eV and above 2.10, 2.10, 2.38, and 2.92~eV, respectively. Finally, the capture of $V$ by $X_s$ is not Coulomb-blocked at any relevant $E_\mathrm{F}$, because $X_s$ is neutral over essentially the whole bandgap.

\begin{table}[htbp]
\centering 
\caption{Fermi-level ranges (in eV, relative to the VBM) for the capture of C$_i$ and $V$ by $XV$. ``Both neutral'': both reactants are neutral, so that the neutral-state binding energies and barriers apply. ``Blocked'': both reactants carry charges of the same sign, so that the capture is suppressed by Coulomb repulsion. In the remaining ranges, only one reactant is charged.}\label{tab:efwindows}
\small
\setlength{\tabcolsep}{4pt}
\begin{tabular}{l c c c c}
  \hline\hline
   & Si & Ge & Sn & Pb \\
  \hline
  \multicolumn{5}{l}{C$_i+XV\rightarrow X_s$} \\
  Both neutral & 0.91--1.50 & 0.91--2.00 & 0.91--2.38 & 1.13--2.92 \\
  Blocked, $E_\mathrm{F}<$ & 0.22 & 0.47 & 0.73 & 0.91 \\
  Blocked, $E_\mathrm{F}>$ & 4.26 & 4.26 & 4.26 & 4.26 \\
  \hline
  \multicolumn{5}{l}{$V+XV\rightarrow XV_2$} \\
  Both neutral & 0.87--1.50 & 0.87--2.00 & 0.87--2.10 & 1.13--2.10 \\
  Blocked, $E_\mathrm{F}<$ & 0.22 & 0.47 & 0.73 & 0.87 \\
  Blocked, $E_\mathrm{F}>$ & 2.10 & 2.10 & 2.38 & 2.92 \\
  \hline\hline
\end{tabular}
\end{table}

Fermi-level control therefore acts very differently on the two conversion channels. The $XV_2$ channel is Coulomb-blocked already in moderately n-type diamond, for $E_\mathrm{F}$ above 2.10--2.92~eV, and in p-type diamond below 0.22--0.87~eV. On the n-type side, by contrast, the capture of C$_i$ is blocked only when C$_i$ itself becomes negatively charged, i.e., for $E_\mathrm{F}$ above 4.26~eV. Nitrogen, with its donor level about 1.7~eV below the CBM, cannot raise $E_\mathrm{F}$ this high, whereas phosphorus, about 0.6~eV below the CBM, can in principle, although at annealing temperatures $E_\mathrm{F}$ is expected to move towards midgap. On the p-type side, the capture of C$_i$ is blocked for $E_\mathrm{F}$ below 0.22--0.91~eV, where both reactants are positively charged. This picture is consistent with charge-state (Coulomb-driven) engineering, which has been demonstrated experimentally to enhance the creation yield and stabilize the negative charge state of color centers---including N$V$, Sn$V$, and Mg$V$---by co-doping with donors such as N, P, O, or S \cite{luhmann_coulomb-driven_2019,luhmann_charge-assisted_2021}. For Sn$V$, the yield of the optically active center after annealing at 1200 $^\circ$C rises from 2.5\% in intrinsic diamond to 8.6\% in phosphorus-doped diamond, while boron doping lowers it to 1.6\% \cite{luhmann_coulomb-driven_2019}. The enhancement with phosphorus is consistent with the blocking of both conversion channels at high $E_\mathrm{F}$. Boron doping would also place $E_\mathrm{F}$ in the range where the capture of C$_i$ by Sn$V$ is Coulomb-blocked ($E_\mathrm{F}<0.73$~eV), yet it lowers the yield of the optically active center. This indicates that the measured optical yield reflects the population of the negative charge state Sn$V^-$ and not only the structural yield; structural probes such as emission channeling on boron-doped samples could separate the two effects. We stress, however, that the experimentally observed benefit of donor doping may equally reflect the stabilization of the negative charge state of the center rather than a suppression of C$_i$ capture, and the Fermi level that maximizes $XV$ formation need not coincide with the one that stabilizes $XV^-$. A recent equilibrium first-principles study likewise found that donor (nitrogen) electrons favor the Sn$V^-$ charge state in n-type diamond \cite{bahulikar_theoretical_2025}. Our results complement this picture by identifying the C$_i$-mediated back-conversion of $XV$ into $X_s$ as an additional channel that Fermi-level control must suppress, and by establishing the trend across the full Si--Pb series.\label{rev:s33end} 

\section{Conclusions}
We systematically studied the formation and stability of $XV$ centers in diamond. The relative formation energies of $XV$, influenced by atomic size, help explain why larger impurities favor $XV$ over substitutional $X_s$. Although the formation of $XV$ centers is energetically favorable, their structural stability and optical activity are influenced by competing defect reactions. The carbon interstitial readily combines with $XV$ to form $X_s$, which is energetically preferred. Formation of $XV_2$ is also thermodynamically favorable, but is kinetically hindered when the vacancy must approach from beyond the second-neighbor shell. These competing pathways may contribute to the low yield of $XV$ centers. The calculated charge transition levels show that oppositely charged reactant pairs never occur, and that Coulomb repulsion blocks the conversions only when both reactants carry charges of the same sign; for the capture of C$_i$, this requires $E_\mathrm{F}$ above 4.26~eV on the n-type side. These results suggest that Fermi-level control may provide a feasible route to tune the competition among these reactions and optimize the formation of group-IV--vacancy color centers in diamond.

\section*{CRediT authorship contribution statement}
\textbf{Xiangru Han:} Investigation, Formal analysis, Visualization, Writing -- original draft. \textbf{Adam Gali:} Conceptualization, Methodology, Supervision, Funding acquisition, Writing -- review \& editing.

\section*{Declaration of Competing Interest}
The authors declare that they have no known competing financial interests or personal relationships that could have appeared to influence the work reported in this paper.

\section*{Data availability}
The data supporting the findings of this study are available from the corresponding author
upon reasonable request.

\section*{Declaration of generative AI and AI-assisted technologies in the manuscript
preparation process}
During the preparation of this work the authors used Claude (Anthropic) in order to check the
internal consistency of the tables and figures, and the values in the main text, and to produce the graphical abstract. 
After using this tool, the authors reviewed and edited the content as needed and take full responsibility for the
content of the published article.

\section*{Acknowledgements}
Funding: This work was supported by the European Union's Horizon Europe research and innovation
programme [grant numbers 101135699 (SPINUS), 101186889 (QuSPARC)]. The funding sources had no
involvement in the study design, in the collection, analysis and interpretation of data, in the
writing of the report, or in the decision to submit the article for publication.

\bibliographystyle{elsarticle-num}
\bibliography{references_checked}

\begin{thebibliography}{10}
\expandafter\ifx\csname url\endcsname\relax
  \def\url#1{\texttt{#1}}\fi
\expandafter\ifx\csname urlprefix\endcsname\relax\def\urlprefix{URL }\fi
\expandafter\ifx\csname href\endcsname\relax
  \def\href#1#2{#2} \def\path#1{#1}\fi

\bibitem{awschalom_quantum_2018}
D.~D. Awschalom, R.~Hanson, J.~Wrachtrup, B.~B. Zhou, Quantum technologies with
  optically interfaced solid-state spins, Nat. Photonics 12 (2018) 516--527.
\newblock \href {https://doi.org/10.1038/s41566-018-0232-2}
  {\path{doi:10.1038/s41566-018-0232-2}}.

\bibitem{atature_material_2018}
M.~Atat\"ure, D.~Englund, N.~Vamivakas, S.-Y. Lee, J.~Wrachtrup, Material
  platforms for spin-based photonic quantum technologies, Nat. Rev. Mater. 3
  (2018) 38--51.
\newblock \href {https://doi.org/10.1038/s41578-018-0008-9}
  {\path{doi:10.1038/s41578-018-0008-9}}.

\bibitem{aharonovich_solid-state_2016}
I.~Aharonovich, D.~Englund, M.~Toth, Solid-state single-photon emitters, Nat.
  Photonics 10 (2016) 631--641.
\newblock \href {https://doi.org/10.1038/nphoton.2016.186}
  {\path{doi:10.1038/nphoton.2016.186}}.

\bibitem{iwasaki_color_2020}
T.~Iwasaki, Color centers based on heavy group-{IV} elements, in: C.~E. Nebel,
  I.~Aharonovich, N.~Mizuochi, M.~Hatano (Eds.), Diamond for {Quantum}
  {Applications} {Part} 1, Vol. 103 of Semiconductors and {Semimetals},
  Elsevier, Amsterdam, 2020, pp. 237--256.
\newblock \href {https://doi.org/10.1016/bs.semsem.2020.03.007}
  {\path{doi:10.1016/bs.semsem.2020.03.007}}.

\bibitem{doherty_nitrogen-vacancy_2013}
M.~W. Doherty, N.~B. Manson, P.~Delaney, F.~Jelezko, J.~Wrachtrup, L.~C.
  Hollenberg, The nitrogen-vacancy colour centre in diamond, Phys. Rep. 528
  (2013) 1--45.
\newblock \href {https://doi.org/10.1016/j.physrep.2013.02.001}
  {\path{doi:10.1016/j.physrep.2013.02.001}}.

\bibitem{abobeih_one-second_2018}
M.~H. Abobeih, J.~Cramer, M.~A. Bakker, N.~Kalb, M.~Markham, D.~J. Twitchen,
  T.~H. Taminiau, One-second coherence for a single electron spin coupled to a
  multi-qubit nuclear-spin environment, Nat. Commun. 9 (2018) 2552.
\newblock \href {https://doi.org/10.1038/s41467-018-04916-z}
  {\path{doi:10.1038/s41467-018-04916-z}}.

\bibitem{thiering_color_2020}
G.~Thiering, A.~Gali, Color centers in diamond for quantum applications, in:
  C.~E. Nebel, I.~Aharonovich, N.~Mizuochi, M.~Hatano (Eds.), Diamond for
  {Quantum} {Applications} {Part} 1, Vol. 103 of Semiconductors and
  {Semimetals}, Elsevier, Amsterdam, 2020, pp. 1--36.
\newblock \href {https://doi.org/10.1016/bs.semsem.2020.03.001}
  {\path{doi:10.1016/bs.semsem.2020.03.001}}.

\bibitem{gruber_scanning_1997}
A.~Gruber, A.~Dr\"abenstedt, C.~Tietz, L.~Fleury, J.~Wrachtrup, C.~{von
  Borczyskowski}, Scanning {Confocal} {Optical} {Microscopy} and {Magnetic}
  {Resonance} on {Single} {Defect} {Centers}, Science 276 (1997) 2012--2014.
\newblock \href {https://doi.org/10.1126/science.276.5321.2012}
  {\path{doi:10.1126/science.276.5321.2012}}.

\bibitem{wrachtrup_quantum_2001}
J.~Wrachtrup, S.~Y. Kilin, A.~P. Nizovtsev, Quantum computation using the
  $^{\textrm{13}}${C} nuclear spins near the single {NV} defect center in
  diamond, Opt. Spectrosc. 91 (2001) 429--437.
\newblock \href {https://doi.org/10.1134/1.1405224}
  {\path{doi:10.1134/1.1405224}}.

\bibitem{gali_ab_2019}
A.~Gali, \textit{Ab initio} theory of the nitrogen-vacancy center in diamond,
  Nanophotonics 8 (2019) 1907--1943.
\newblock \href {https://doi.org/10.1515/nanoph-2019-0154}
  {\path{doi:10.1515/nanoph-2019-0154}}.

\bibitem{waldherr_dark_2011}
G.~Waldherr, J.~Beck, M.~Steiner, P.~Neumann, A.~Gali, T.~Frauenheim,
  F.~Jelezko, J.~Wrachtrup, Dark {States} of {Single} {Nitrogen}-{Vacancy}
  {Centers} in {Diamond} {Unraveled} by {Single} {Shot} {NMR}, Phys. Rev. Lett.
  106 (2011) 157601.
\newblock \href {https://doi.org/10.1103/PhysRevLett.106.157601}
  {\path{doi:10.1103/PhysRevLett.106.157601}}.

\bibitem{beha_optimum_2012}
K.~Beha, A.~Batalov, N.~B. Manson, R.~Bratschitsch, A.~Leitenstorfer, Optimum
  {Photoluminescence} {Excitation} and {Recharging} {Cycle} of {Single}
  {Nitrogen}-{Vacancy} {Centers} in {Ultrapure} {Diamond}, Phys. Rev. Lett. 109
  (2012) 097404.
\newblock \href {https://doi.org/10.1103/PhysRevLett.109.097404}
  {\path{doi:10.1103/PhysRevLett.109.097404}}.

\bibitem{casola_probing_2018}
F.~Casola, T.~van~der Sar, A.~Yacoby, Probing condensed matter physics with
  magnetometry based on nitrogen-vacancy centres in diamond, Nat. Rev. Mater. 3
  (2018) 17088.
\newblock \href {https://doi.org/10.1038/natrevmats.2017.88}
  {\path{doi:10.1038/natrevmats.2017.88}}.

\bibitem{norman_novel_2021}
V.~A. Norman, S.~Majety, Z.~Wang, W.~H. Casey, N.~Curro, M.~Radulaski, Novel
  color center platforms enabling fundamental scientific discovery, InfoMat 3
  (2021) 869--890.
\newblock \href {https://doi.org/10.1002/inf2.12128}
  {\path{doi:10.1002/inf2.12128}}.

\bibitem{gaebel_room-temperature_2006}
T.~Gaebel, M.~Domhan, I.~Popa, C.~Wittmann, P.~Neumann, F.~Jelezko, J.~R.
  Rabeau, N.~Stavrias, A.~D. Greentree, S.~Prawer, J.~Meijer, J.~Twamley, P.~R.
  Hemmer, J.~Wrachtrup, Room-temperature coherent coupling of single spins in
  diamond, Nat. Phys. 2 (2006) 408--413.
\newblock \href {https://doi.org/10.1038/nphys318}
  {\path{doi:10.1038/nphys318}}.

\bibitem{santori_nanophotonics_2010}
C.~Santori, P.~E. Barclay, K.-M.~C. Fu, R.~G. Beausoleil, S.~Spillane,
  M.~Fisch, Nanophotonics for quantum optics using nitrogen-vacancy centers in
  diamond, Nanotechnology 21 (2010) 274008.
\newblock \href {https://doi.org/10.1088/0957-4484/21/27/274008}
  {\path{doi:10.1088/0957-4484/21/27/274008}}.

\bibitem{hepp_electronic_2014}
C.~Hepp, T.~M\"uller, V.~Waselowski, J.~N. Becker, B.~Pingault,
  H.~Sternschulte, D.~Steinm\"uller-Nethl, A.~Gali, J.~R. Maze, M.~Atat\"ure,
  C.~Becher, Electronic {Structure} of the {Silicon} {Vacancy} {Color} {Center}
  in {Diamond}, Phys. Rev. Lett. 112 (2014) 036405.
\newblock \href {https://doi.org/10.1103/PhysRevLett.112.036405}
  {\path{doi:10.1103/PhysRevLett.112.036405}}.

\bibitem{siyushev_optical_2017}
P.~Siyushev, M.~H. Metsch, A.~Ijaz, J.~M. Binder, M.~K. Bhaskar, D.~D.
  Sukachev, A.~Sipahigil, R.~E. Evans, C.~T. Nguyen, M.~D. Lukin, P.~R. Hemmer,
  Y.~N. Palyanov, I.~N. Kupriyanov, Y.~M. Borzdov, L.~J. Rogers, F.~Jelezko,
  Optical and microwave control of germanium-vacancy center spins in diamond,
  Phys. Rev. B 96 (2017) 081201.
\newblock \href {https://doi.org/10.1103/PhysRevB.96.081201}
  {\path{doi:10.1103/PhysRevB.96.081201}}.

\bibitem{rugar_quantum_2021}
A.~E. Rugar, S.~Aghaeimeibodi, D.~Riedel, C.~Dory, H.~Lu, P.~J. McQuade, Z.-X.
  Shen, N.~A. Melosh, J.~Vu\v{c}kovi\'c, Quantum {Photonic} {Interface} for
  {Tin}-{Vacancy} {Centers} in {Diamond}, Phys. Rev. X 11 (2021) 031021.
\newblock \href {https://doi.org/10.1103/PhysRevX.11.031021}
  {\path{doi:10.1103/PhysRevX.11.031021}}.

\bibitem{trusheim_lead-related_2019}
M.~E. Trusheim, N.~H. Wan, K.~C. Chen, C.~J. Ciccarino, J.~Flick,
  R.~Sundararaman, G.~Malladi, E.~Bersin, M.~Walsh, B.~Lienhard, H.~Bakhru,
  P.~Narang, D.~Englund, Lead-related quantum emitters in diamond, Phys. Rev. B
  99 (2019) 075430.
\newblock \href {https://doi.org/10.1103/PhysRevB.99.075430}
  {\path{doi:10.1103/PhysRevB.99.075430}}.

\bibitem{thiering_ab_2018}
G.~Thiering, A.~Gali, \textit{{Ab} {Initio}} {Magneto}-{Optical} {Spectrum} of
  {Group}-{IV} {Vacancy} {Color} {Centers} in {Diamond}, Phys. Rev. X 8 (2018)
  021063.
\newblock \href {https://doi.org/10.1103/PhysRevX.8.021063}
  {\path{doi:10.1103/PhysRevX.8.021063}}.

\bibitem{chen_building_2020}
D.~Chen, N.~Zheludev, W.-b. Gao, Building {Blocks} for {Quantum} {Network}
  {Based} on {Group}-{IV} {Split}-{Vacancy} {Centers} in {Diamond}, Adv.
  Quantum Technol. 3 (2020) 1900069.
\newblock \href {https://doi.org/10.1002/qute.201900069}
  {\path{doi:10.1002/qute.201900069}}.

\bibitem{arjona_martinez_photonic_2022}
J.~Arjona~Mart\'inez, R.~A. Parker, K.~C. Chen, C.~M. Purser, L.~Li, C.~P.
  Michaels, A.~M. Stramma, R.~Debroux, I.~B. Harris, M.~Hayhurst~Appel, E.~C.
  Nichols, M.~E. Trusheim, D.~A. Gangloff, D.~Englund, M.~Atat\"ure, Photonic
  {Indistinguishability} of the {Tin}-{Vacancy} {Center} in {Nanostructured}
  {Diamond}, Phys. Rev. Lett. 129 (2022) 173603.
\newblock \href {https://doi.org/10.1103/PhysRevLett.129.173603}
  {\path{doi:10.1103/PhysRevLett.129.173603}}.

\bibitem{bradac_quantum_2019}
C.~Bradac, W.~Gao, J.~Forneris, M.~E. Trusheim, I.~Aharonovich, Quantum
  nanophotonics with group {IV} defects in diamond, Nat. Commun. 10 (2019)
  5625.
\newblock \href {https://doi.org/10.1038/s41467-019-13332-w}
  {\path{doi:10.1038/s41467-019-13332-w}}.

\bibitem{schroder_scalable_2017}
T.~Schr\"oder, M.~E. Trusheim, M.~Walsh, L.~Li, J.~Zheng, M.~Schukraft,
  A.~Sipahigil, R.~E. Evans, D.~D. Sukachev, C.~T. Nguyen, J.~L. Pacheco, R.~M.
  Camacho, E.~S. Bielejec, M.~D. Lukin, D.~Englund, Scalable focused ion beam
  creation of nearly lifetime-limited single quantum emitters in diamond
  nanostructures, Nat. Commun. 8 (2017) 15376.
\newblock \href {https://doi.org/10.1038/ncomms15376}
  {\path{doi:10.1038/ncomms15376}}.

\bibitem{ju_preparations_2021}
Z.~Ju, J.~Lin, S.~Shen, B.~Wu, E.~Wu, Preparations and applications of single
  color centers in diamond, Adv. Phys. X 6 (2021) 1858721.
\newblock \href {https://doi.org/10.1080/23746149.2020.1858721}
  {\path{doi:10.1080/23746149.2020.1858721}}.

\bibitem{wang_single_2006}
C.~Wang, C.~Kurtsiefer, H.~Weinfurter, B.~Burchard, Single photon emission from
  {SiV} centres in diamond produced by ion implantation, J. Phys. B: At. Mol.
  Opt. Phys. 39 (2006) 37--41.
\newblock \href {https://doi.org/10.1088/0953-4075/39/1/005}
  {\path{doi:10.1088/0953-4075/39/1/005}}.

\bibitem{neu_single_2011}
E.~Neu, D.~Steinmetz, J.~Riedrich-M\"oller, S.~Gsell, M.~Fischer, M.~Schreck,
  C.~Becher, Single photon emission from silicon-vacancy colour centres in
  chemical vapour deposition nano-diamonds on iridium, New J. Phys. 13 (2011)
  025012.
\newblock \href {https://doi.org/10.1088/1367-2630/13/2/025012}
  {\path{doi:10.1088/1367-2630/13/2/025012}}.

\bibitem{iwasaki_germanium-vacancy_2015}
T.~Iwasaki, F.~Ishibashi, Y.~Miyamoto, Y.~Doi, S.~Kobayashi, T.~Miyazaki,
  K.~Tahara, K.~D. Jahnke, L.~J. Rogers, B.~Naydenov, F.~Jelezko, S.~Yamasaki,
  S.~Nagamachi, T.~Inubushi, N.~Mizuochi, M.~Hatano, Germanium-{Vacancy}
  {Single} {Color} {Centers} in {Diamond}, Sci. Rep. 5 (2015) 12882.
\newblock \href {https://doi.org/10.1038/srep12882}
  {\path{doi:10.1038/srep12882}}.

\bibitem{tchernij_single-photon-emitting_2017}
S.~D. Tchernij, T.~Herzig, J.~Forneris, J.~K\"upper, S.~Pezzagna, P.~Traina,
  E.~Moreva, I.~P. Degiovanni, G.~Brida, N.~Skukan, M.~Genovese,
  M.~Jak\v{s}i\'c, J.~Meijer, P.~Olivero, Single-{Photon}-{Emitting} {Optical}
  {Centers} in {Diamond} {Fabricated} upon {Sn} {Implantation}, ACS Photonics 4
  (2017) 2580--2586.
\newblock \href {https://doi.org/10.1021/acsphotonics.7b00904}
  {\path{doi:10.1021/acsphotonics.7b00904}}.

\bibitem{iwasaki_tin-vacancy_2017}
T.~Iwasaki, Y.~Miyamoto, T.~Taniguchi, P.~Siyushev, M.~H. Metsch, F.~Jelezko,
  M.~Hatano, Tin-{Vacancy} {Quantum} {Emitters} in {Diamond}, Phys. Rev. Lett.
  119~(25) (2017) 253601.
\newblock \href {https://doi.org/10.1103/PhysRevLett.119.253601}
  {\path{doi:10.1103/PhysRevLett.119.253601}}.

\bibitem{ditalia_tchernij_single-photon_2018}
S.~Ditalia~Tchernij, T.~L\"uhmann, T.~Herzig, J.~K\"upper, A.~Damin,
  S.~Santonocito, M.~Signorile, P.~Traina, E.~Moreva, F.~Celegato, S.~Pezzagna,
  I.~P. Degiovanni, P.~Olivero, M.~Jak\v{s}i\'c, J.~Meijer, M.~Genovese,
  J.~Forneris, Single-{Photon} {Emitters} in {Lead}-{Implanted}
  {Single}-{Crystal} {Diamond}, ACS Photonics 5~(12) (2018) 4864--4871.
\newblock \href {https://doi.org/10.1021/acsphotonics.8b01013}
  {\path{doi:10.1021/acsphotonics.8b01013}}.

\bibitem{wahl_structural_2024}
U.~Wahl, J.~a.~G. Correia, A.~Costa, A.~Lamelas, V.~Amaral, K.~Johnston,
  G.~Magchiels, S.~M. Tunhuma, A.~Vantomme, L.~M.~C. Pereira, Structural
  formation yield of {GeV} centers from implanted {Ge} in diamond, Mater.
  Quantum Technol. 4 (2024) 025101.
\newblock \href {https://doi.org/10.1088/2633-4356/ad4b8d}
  {\path{doi:10.1088/2633-4356/ad4b8d}}.

\bibitem{wahl_direct_2020}
U.~Wahl, J.~G. Correia, R.~Villarreal, E.~Bourgeois, M.~Gulka, M.~Nesl\'adek,
  A.~Vantomme, L.~M.~C. Pereira, Direct {Structural} {Identification} and
  {Quantification} of the {Split}-{Vacancy} {Configuration} for {Implanted}
  {Sn} in {Diamond}, Phys. Rev. Lett. 125~(4) (2020) 045301.
\newblock \href {https://doi.org/10.1103/PhysRevLett.125.045301}
  {\path{doi:10.1103/PhysRevLett.125.045301}}.

\bibitem{cheng_laser_2025}
X.~Cheng, A.~Thurn, G.~Chen, G.~S. Jones, J.~E. Bennett, M.~Coke, M.~Adshead,
  C.~P. Michaels, O.~Balci, A.~C. Ferrari, M.~Atat\"ure, R.~J. Curry, J.~M.
  Smith, P.~S. Salter, D.~A. Gangloff, Laser activation of single group-{IV}
  colour centres in diamond, Nat. Commun. 16 (2025) 5124.
\newblock \href {https://doi.org/10.1038/s41467-025-60373-5}
  {\path{doi:10.1038/s41467-025-60373-5}}.

\bibitem{pugliese_photoactivation_2025}
V.~Pugliese, E.~Nieto~Hern\'andez, E.~Corte, M.~Govoni, S.~Ditalia~Tchernij,
  P.~Olivero, J.~Forneris, Photoactivation of {Color} {Centers} {Induced} by
  {CW} {Laser} {Irradiation} in {Ion}-{Implanted} {Diamond}, ACS Photonics 12
  (2025) 3803--3814.
\newblock \href {https://doi.org/10.1021/acsphotonics.5c00826}
  {\path{doi:10.1021/acsphotonics.5c00826}}.

\bibitem{fu_molecular_2019}
X.~Fu, Z.~Xu, Z.~He, A.~Hartmaier, F.~Fang, Molecular dynamics simulation of
  silicon ion implantation into diamond and subsequent annealing, Nucl.
  Instrum. Methods Phys. Res. B 450 (2019) 51--55.
\newblock \href {https://doi.org/10.1016/j.nimb.2018.04.027}
  {\path{doi:10.1016/j.nimb.2018.04.027}}.

\bibitem{kuate_defo_how_2019}
R.~Kuate~Defo, E.~Kaxiras, S.~L. Richardson, How carbon vacancies can affect
  the properties of group {IV} color centers in diamond: {A} study of
  thermodynamics and kinetics, J. Appl. Phys. 126 (2019) 195103.
\newblock \href {https://doi.org/10.1063/1.5123227}
  {\path{doi:10.1063/1.5123227}}.

\bibitem{chen_simulating_2025}
G.~Chen, J.~C.~A. Prentice, J.~M. Smith, Simulating the dynamics of {NV$^-$}
  formation in diamond in the presence of carbon self-interstitials, npj
  Comput. Mater. 11 (2025) 157.
\newblock \href {https://doi.org/10.1038/s41524-025-01605-6}
  {\path{doi:10.1038/s41524-025-01605-6}}.

\bibitem{kresse_efficient_1996}
G.~Kresse, J.~Furthm\"uller, Efficient iterative schemes for \textit{ab initio}
  total-energy calculations using a plane-wave basis set, Phys. Rev. B 54
  (1996) 11169--11186.
\newblock \href {https://doi.org/10.1103/PhysRevB.54.11169}
  {\path{doi:10.1103/PhysRevB.54.11169}}.

\bibitem{blochl_projector_1994}
P.~E. Bl\"ochl, Projector augmented-wave method, Phys. Rev. B 50 (1994)
  17953--17979.
\newblock \href {https://doi.org/10.1103/PhysRevB.50.17953}
  {\path{doi:10.1103/PhysRevB.50.17953}}.

\bibitem{kresse_ultrasoft_1999}
G.~Kresse, D.~Joubert, From ultrasoft pseudopotentials to the projector
  augmented-wave method, Phys. Rev. B 59~(3) (1999) 1758--1775.
\newblock \href {https://doi.org/10.1103/PhysRevB.59.1758}
  {\path{doi:10.1103/PhysRevB.59.1758}}.

\bibitem{heyd_hybrid_2003}
J.~Heyd, G.~E. Scuseria, M.~Ernzerhof, Hybrid functionals based on a screened
  {Coulomb} potential, J. Chem. Phys. 118 (2003) 8207--8215.
\newblock \href {https://doi.org/10.1063/1.1564060}
  {\path{doi:10.1063/1.1564060}}.

\bibitem{heyd_erratum_2006}
J.~Heyd, G.~E. Scuseria, M.~Ernzerhof, Erratum: {\textquotedblleft}{Hybrid}
  functionals based on a screened {Coulomb} potential{\textquotedblright} [{J}.
  {Chem}. {Phys}. \textbf{118}, 8207 (2003)], J. Chem. Phys. 124~(21) (2006)
  219906.
\newblock \href {https://doi.org/10.1063/1.2204597}
  {\path{doi:10.1063/1.2204597}}.

\bibitem{krukau_influence_2006}
A.~V. Krukau, O.~A. Vydrov, A.~F. Izmaylov, G.~E. Scuseria, Influence of the
  exchange screening parameter on the performance of screened hybrid
  functionals, J. Chem. Phys. 125~(22) (2006) 224106.
\newblock \href {https://doi.org/10.1063/1.2404663}
  {\path{doi:10.1063/1.2404663}}.

\bibitem{henkelman_climbing_2000}
G.~Henkelman, B.~P. Uberuaga, H.~J\'onsson, A climbing image nudged elastic
  band method for finding saddle points and minimum energy paths, J. Chem.
  Phys. 113 (2000) 9901--9904.
\newblock \href {https://doi.org/10.1063/1.1329672}
  {\path{doi:10.1063/1.1329672}}.

\bibitem{perdew_generalized_1996}
J.~P. Perdew, K.~Burke, M.~Ernzerhof, Generalized {Gradient} {Approximation}
  {Made} {Simple}, Phys. Rev. Lett. 77 (1996) 3865--3868.
\newblock \href {https://doi.org/10.1103/PhysRevLett.77.3865}
  {\path{doi:10.1103/PhysRevLett.77.3865}}.

\bibitem{deak_formation_2014}
P.~De\'ak, B.~Aradi, M.~Kaviani, T.~Frauenheim, A.~Gali, Formation of {NV}
  centers in diamond: {A} theoretical study based on calculated transitions and
  migration of nitrogen and vacancy related defects, Phys. Rev. B 89 (2014)
  075203.
\newblock \href {https://doi.org/10.1103/PhysRevB.89.075203}
  {\path{doi:10.1103/PhysRevB.89.075203}}.

\bibitem{breuer_ab_1995}
S.~J. Breuer, P.~R. Briddon, Ab initio investigation of the native defects in
  diamond and self-diffusion, Phys. Rev. B 51 (1995) 6984.
\newblock \href {https://doi.org/10.1103/PhysRevB.51.6984}
  {\path{doi:10.1103/PhysRevB.51.6984}}.

\bibitem{zhang_chemical_1991}
S.~B. Zhang, J.~E. Northrup, Chemical potential dependence of defect formation
  energies in {GaAs}: {Application} to {Ga} self-diffusion, Phys. Rev. Lett. 67
  (1991) 2339--2342.
\newblock \href {https://doi.org/10.1103/PhysRevLett.67.2339}
  {\path{doi:10.1103/PhysRevLett.67.2339}}.

\bibitem{freysoldt_first-principles_2014}
C.~Freysoldt, B.~Grabowski, T.~Hickel, J.~Neugebauer, G.~Kresse, A.~Janotti,
  C.~G. Van~de Walle, First-principles calculations for point defects in
  solids, Rev. Mod. Phys. 86 (2014) 253--305.
\newblock \href {https://doi.org/10.1103/RevModPhys.86.253}
  {\path{doi:10.1103/RevModPhys.86.253}}.

\bibitem{freysoldt_fully_2009}
C.~Freysoldt, J.~Neugebauer, C.~G. Van~de Walle, Fully \textit{{Ab} {Initio}}
  {Finite}-{Size} {Corrections} for {Charged}-{Defect} {Supercell}
  {Calculations}, Phys. Rev. Lett. 102 (2009) 016402.
\newblock \href {https://doi.org/10.1103/PhysRevLett.102.016402}
  {\path{doi:10.1103/PhysRevLett.102.016402}}.

\bibitem{iakoubovskii_annealing_2003}
K.~Iakoubovskii, I.~Kiflawi, K.~Johnston, A.~Collins, G.~Davies, A.~Stesmans,
  Annealing of vacancies and interstitials in diamond, Physica B 340-342 (2003)
  67--75.
\newblock \href {https://doi.org/10.1016/j.physb.2003.09.005}
  {\path{doi:10.1016/j.physb.2003.09.005}}.

\bibitem{allers_annealing_1998}
L.~Allers, A.~T. Collins, J.~Hiscock, The annealing of interstitial-related
  optical centres in type {II} natural and {CVD} diamond, Diam. Relat. Mater. 7
  (1998) 228--232.
\newblock \href {https://doi.org/10.1016/S0925-9635(97)00161-1}
  {\path{doi:10.1016/S0925-9635(97)00161-1}}.

\bibitem{goss_vacancy-impurity_2005}
J.~P. Goss, P.~R. Briddon, M.~J. Rayson, S.~J. Sque, R.~Jones, Vacancy-impurity
  complexes and limitations for implantation doping of diamond, Phys. Rev. B 72
  (2005) 035214.
\newblock \href {https://doi.org/10.1103/PhysRevB.72.035214}
  {\path{doi:10.1103/PhysRevB.72.035214}}.

\bibitem{mainwood_nitrogen_1994}
A.~Mainwood, Nitrogen and nitrogen-vacancy complexes and their formation in
  diamond, Phys. Rev. B 49 (1994) 7934--7940.
\newblock \href {https://doi.org/10.1103/PhysRevB.49.7934}
  {\path{doi:10.1103/PhysRevB.49.7934}}.

\bibitem{naydenov_increasing_2010}
B.~Naydenov, F.~Reinhard, A.~L\"ammle, V.~Richter, R.~Kalish, U.~F.~S.
  D'Haenens-Johansson, M.~Newton, F.~Jelezko, J.~Wrachtrup, Increasing the
  coherence time of single electron spins in diamond by high temperature
  annealing, Appl. Phys. Lett. 97 (2010) 242511.
\newblock \href {https://doi.org/10.1063/1.3527975}
  {\path{doi:10.1063/1.3527975}}.

\bibitem{pezzagna_creation_2010}
S.~Pezzagna, B.~Naydenov, F.~Jelezko, J.~Wrachtrup, J.~Meijer, Creation
  efficiency of nitrogen-vacancy centres in diamond, New J. Phys. 12 (2010)
  065017.
\newblock \href {https://doi.org/10.1088/1367-2630/12/6/065017}
  {\path{doi:10.1088/1367-2630/12/6/065017}}.

\bibitem{rabeau_implantation_2006}
J.~R. Rabeau, P.~Reichart, G.~Tamanyan, D.~N. Jamieson, S.~Prawer, F.~Jelezko,
  T.~Gaebel, I.~Popa, M.~Domhan, J.~Wrachtrup, Implantation of labelled single
  nitrogen vacancy centers in diamond using $^{\textrm{15}}${N}, Appl. Phys.
  Lett. 88 (2006) 023113.
\newblock \href {https://doi.org/10.1063/1.2158700}
  {\path{doi:10.1063/1.2158700}}.

\bibitem{baba_annealing_2025}
T.~Baba, M.~Iizawa, K.~Takenaka, K.~Kimura, A.~Kawasaki, T.~Taniguchi,
  M.~Miyakawa, H.~Okazaki, O.~Hanaizumi, S.~Onoda, Annealing time dependence on
  creation of {SiV}, {GeV}, and {SnV} in diamond by atmospheric annealing at
  1800\,$^{\circ}$c, Phys. Status Solidi A 222~(5) (2025) 2400303.
\newblock \href {https://doi.org/10.1002/pssa.202400303}
  {\path{doi:10.1002/pssa.202400303}}.

\bibitem{tzeng_improving_2024}
Y.-K. Tzeng, F.~Ke, C.~Jia, Y.~Liu, S.~Park, M.~Han, M.~Frost, X.~Cai, W.~L.
  Mao, R.~C. Ewing, Y.~Cui, T.~P. Devereaux, Y.~Lin, S.~Chu, Improving the
  creation of {SiV} centers in diamond via sub-{\textmu}s pulsed annealing
  treatment, Nat. Commun. 15 (2024) 7251.
\newblock \href {https://doi.org/10.1038/s41467-024-51523-2}
  {\path{doi:10.1038/s41467-024-51523-2}}.

\bibitem{goss_self-interstitial_2001}
J.~P. Goss, B.~J. Coomer, R.~Jones, T.~D. Shaw, P.~R. Briddon, M.~Rayson,
  S.~\"Oberg, Self-interstitial aggregation in diamond, Phys. Rev. B 63 (2001)
  195208.
\newblock \href {https://doi.org/10.1103/PhysRevB.63.195208}
  {\path{doi:10.1103/PhysRevB.63.195208}}.

\bibitem{goss_density_2007}
J.~P. Goss, P.~R. Briddon, M.~J. Shaw, Density functional simulations of
  silicon-containing point defects in diamond, Phys. Rev. B 76 (2007) 075204.
\newblock \href {https://doi.org/10.1103/PhysRevB.76.075204}
  {\path{doi:10.1103/PhysRevB.76.075204}}.

\bibitem{thiering_complexes_2015}
G.~Thiering, A.~Gali, Complexes of silicon, vacancy, and hydrogen in diamond:
  {A} density functional study, Phys. Rev. B 92 (2015) 165203.
\newblock \href {https://doi.org/10.1103/PhysRevB.92.165203}
  {\path{doi:10.1103/PhysRevB.92.165203}}.

\bibitem{dhaenens-johansson_epr_2010}
U.~F.~S. D'Haenens-Johansson, A.~M. Edmonds, M.~E. Newton, J.~P. Goss, P.~R.
  Briddon, J.~M. Baker, P.~M. Martineau, R.~U.~A. Khan, D.~J. Twitchen, S.~D.
  Williams, {EPR} of a defect in {CVD} diamond involving both silicon and
  hydrogen that shows preferential alignment, Phys. Rev. B 82 (2010) 155205.
\newblock \href {https://doi.org/10.1103/PhysRevB.82.155205}
  {\path{doi:10.1103/PhysRevB.82.155205}}.

\bibitem{davies_vacancy-related_1992}
G.~Davies, S.~C. Lawson, A.~T. Collins, A.~Mainwood, S.~J. Sharp,
  Vacancy-related centers in diamond, Phys. Rev. B 46~(20) (1992) 13157--13170.
\newblock \href {https://doi.org/10.1103/PhysRevB.46.13157}
  {\path{doi:10.1103/PhysRevB.46.13157}}.

\bibitem{luhmann_coulomb-driven_2019}
T.~L\"uhmann, R.~John, R.~Wunderlich, J.~Meijer, S.~Pezzagna, Coulomb-driven
  single defect engineering for scalable qubits and spin sensors in diamond,
  Nat. Commun. 10 (2019) 4956.
\newblock \href {https://doi.org/10.1038/s41467-019-12556-0}
  {\path{doi:10.1038/s41467-019-12556-0}}.

\bibitem{luhmann_charge-assisted_2021}
T.~L\"uhmann, J.~Meijer, S.~Pezzagna, Charge-{Assisted} {Engineering} of
  {Color} {Centers} in {Diamond}, Phys. Status Solidi A 218~(5) (2021) 2000614.
\newblock \href {https://doi.org/10.1002/pssa.202000614}
  {\path{doi:10.1002/pssa.202000614}}.

\bibitem{bahulikar_theoretical_2025}
A.~Bahulikar, S.~L. Richardson, R.~Kuate~Defo, Theoretical investigation of
  preparation pathways that enhance the equilibrium yield of negatively ionized
  tin-vacancy centers in {N}-doped diamond, J. Appl. Phys. 137~(17) (2025)
  175110.
\newblock \href {https://doi.org/10.1063/5.0261260}
  {\path{doi:10.1063/5.0261260}}.

\end{thebibliography}

\end{document}